\documentclass[twocolumn,superscriptaddress]{revtex4-1}
\usepackage{graphicx}
\usepackage{color}

\begin{document}

\title{{\bf Size effect in the ionization energy of PAH clusters}}

\author{C. Joblin}
\email[Corresponding author:~]{christine.joblin@irap.omp.eu}
\affiliation{Universit\'e de Toulouse, UPS-OMP, IRAP, Toulouse, France}
\affiliation{CNRS, IRAP, 9 Av. colonel Roche, BP 44346, 31028, Toulouse Cedex 4, France}
\author{L. Dontot}
\affiliation{Universit\'e de Toulouse, UPS-OMP, IRAP, Toulouse, France}
\affiliation{CNRS, IRAP, 9 Av. colonel Roche, BP 44346, 31028, Toulouse Cedex 4, France}
\author{G.A. Garcia}
\affiliation{Synchrotron SOLEIL, L'Orme des Merisiers, 91192 Gif sur Yvette Cedex, France}
\author{F. Spiegelman}
\affiliation{Laboratoire de Chimie et Physique Quantiques LCPQ/IRSAMC, Universit\'e de Toulouse (UPS) and CNRS,
118 Route de Narbonne, 31062 Toulouse, France}
\author{M. Rapacioli}
\affiliation{Laboratoire de Chimie et Physique Quantiques LCPQ/IRSAMC, Universit\'e de Toulouse (UPS) and CNRS,
118 Route de Narbonne, 31062 Toulouse, France}
\author{L. Nahon}
\affiliation{Synchrotron SOLEIL, L'Orme des Merisiers, 91192 Gif sur Yvette Cedex, France}
\author{P. Parneix}
\affiliation{Institut des Sciences Mol\'eculaires d'Orsay, CNRS, Univ Paris Sud, Universit\'e Paris-Saclay, F-91405 Orsay, France}
\author{T. Pino}
\affiliation{Institut des Sciences Mol\'eculaires d'Orsay, CNRS, Univ Paris Sud, Universit\'e Paris-Saclay, F-91405 Orsay, France}
\author{P. Br\'echignac}
\affiliation{Institut des Sciences Mol\'eculaires d'Orsay, CNRS, Univ Paris Sud, Universit\'e Paris-Saclay, F-91405 Orsay, France}

\date{\today}

\begin{abstract}
We report the first experimental measurement of the near-threshold photo-ionization spectra of polycyclic aromatic hydrocarbon clusters made of pyrene C$_{16}$H$_{10}$ and coronene C$_{24}$H$_{12}$, obtained using imaging photoelectron � photoion coincidence spectrometry with a VUV synchrotron beamline. The experimental results of the ionization energy are confronted to calculated ones obtained from simulations using dedicated electronic structure treatment for large ionized molecular clusters. Experiment and theory consistently find a decrease of the ionization energy with cluster size. The inclusion of temperature effects in the simulations leads to a lowering of this energy and to a quantitative agreement with the experiment. In the case of pyrene, both theory and experiment show a discontinuity in the IE trend for the hexamer.
\end{abstract}

\maketitle

For almost 30 years, astronomers, physicists and chemists have striven to assess the astrophysical significance of interstellar polycyclic aromatic hydrocarbons (PAHs) (see Ref.\,\cite{PAHUniverse} for a recent compilation of articles). The presence of these PAHs has been proposed to account for a set of aromatic infrared bands in the 3-15 $\mu$m range. These bands are observed in emission from interstellar and circumstellar regions that are exposed to ultraviolet stellar light, which are generally referred to as photodissociation regions (PDRs). However, the main drawback of the initial PAH hypothesis remains, i.e. that no individual species have been identified so far.  Still, the detection of C$_{60}$ and C$_{60}^+$ through their infrared signatures has recently shown that large gas-phase aromatic molecules exist in PDRs \cite{sellgren10, berne13}. Rather than looking for the spectral signatures of specific molecules, one could constrain the chemical scenarios that lead to the formation of these large carbonaceous molecules in space. In PDRs, the analysis of astronomical observations suggests that these species are produced by photo-evaporation of nanograins, which molecular clusters could mimic \cite{rapacioli05}.  A few studies have been performed to evaluate the formation and survival of neutral PAH clusters in PDRs \cite{rapacioli06, montillaud14}. Ionized PAH clusters have also been considered as species of astrophysical interest since their binding energies are expected to be larger than the corresponding neutrals and therefore they are expected to survive on a longer timescale in PDRs \cite{rapacioli06, rapacioli09}. The astrophysical application has motivated experimental studies on the interaction of PAH and/or C$_{60}$ clusters with UV laser photons (h$\nu$= 4\,eV) \cite{brechignac05} and with fast ions beams  \cite{holm10, zettergren13, gatchell14, delaunay15}. In addition to the study of the induced fragmentation routes for such clusters, these works demonstrated the possibility for molecular growth inside the clusters.
In the modeling studies mentioned above, the properties of PAH clusters have been simulated since quantitive data from experiments is still rather limited. Yet, describing the electronic structure of large clusters with up to a 1000 atoms remains a challenging task for theory. Describing charge resonance in ionized clusters constitutes another difficulty \cite{rapacioli09}. Experimental data is therefore critical to benchmark models. In the present study, we report  the first experimental measurement of spectral features accompanying the ionization onset of PAH clusters made of pyrene C$_{16}$H$_{10}$ and coronene C$_{24}$H$_{12}$. The experimental results are confronted to those obtained from theoretical simulations at 0\,K and at finite temperatures. This yields important information about the interplay between ionization energies and geometries of these clusters.

\begin{figure}[ht]
 \begin{center}
  \includegraphics[width=8.5cm]{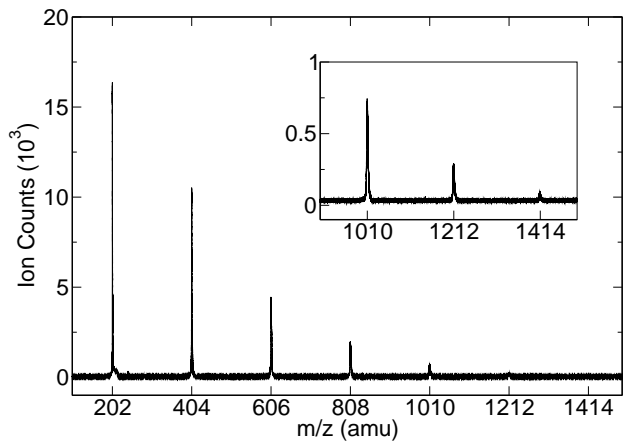}
 \end{center}
\caption{Mass spectrum showing the pyrene cluster distribution, from the time-of-flight spectrometer integrated over all the photon energies measured in one of the performed scans.}\label{fig_tof}
\end{figure}

Experiments have been performed at the DESIRS VUV beamline \cite{Nahon12} of the French synchrotron facility SOLEIL using the molecular beam setup available at the SAPHIRS permanent endstation coupled to an imaging photoelectron � photoion coincidence (iPEPICO)  spectrometry technique. Due to the low vapour pressure of PAHs, an in-vacuum stainless-steel oven especially designed to reach high temperatures (up to ~500 $^\circ$C) was built. Typically 1.5 gr of coronene (99\% purity from Sigma-Aldrich) or  pyrene (98\% purity from Fluka) were introduced into the oven and a working temperature of 340-360\,$^\circ$C and 180\,$^\circ$C was used for coronene and pyrene, respectively. The PAH vapour was mixed and driven by an Ar flow, then expanded through a 50\,$\mu$m nozzle; the backing pressure, 1.1 to 1.2 bars,  was adjusted to maximize the production of PAH clusters. The maximum size obtained for coronene clusters (up to the pentamer) and pyrene clusters (up to the heptamer) resulted from our choice to maintain stable expansion conditions in the cluster source while scanning the photon energy. 
The free jet was skimmed through a 2-mm-ID orifice,  and passed through the ionization zone, while crossing the VUV synchrotron beam at right angle, at the centre of the iPEPICO spectrometer. We used DELICIOUS 2 \cite{garcia09} for the coronene experiments reported here and its upgraded version DELICIOUS 3 \cite{garcia13} for the pyrene clusters. Briefly, DELICIOUS 2 consists of a velocity map imaging (VMI) analyzer, coupled in coincidence to a Wiley-McLaren time of flight (WM-TOF) spectrometer, which is able to provide mass-filtered photoelectron or photoion images. The ion images were recorded at several photon energies only to ensure that the detected clusters had no measurable translational energy within the photon energy range presented in this work, and thus do not originate from a dissociative ionization event, but rather from direct ionization of the neutral counterpart. 
DELICIOUS 3 has the ability to provide photoelectron and photoion images simultaneously, which means that on top of the mass selection,  ion translational energy selection is performed, again to consider only the parent ions, and discard any possible fragments.
Since not all the capabilities of the newer spectrometer were used, the only real difference between both set of experiments is the improved mass resolution for the pyrene runs, which did not affect the relative quality of the results.
In all the experiments, the electrostatic configuration was set as to ensure full transmission for all ions and electrons. The decrease of signal for increasing size observed in Fig.\,\ref{fig_tof} is only attributed to the clusters relative abundance in the molecular beam, and to the efficiency decrease of the microchannel plate ion detectors, estimated at 30\% between 300 and 1500 amu.
The cation spectroscopy was recorded with the threshold photoelectron technique by scanning the photon energy and taking into account only the slow electrons, as previously described \cite{poully10}. 

\begin{figure*}[ht]
 \begin{center}
  \begin{tabular}{cc}
   \includegraphics[width=8.5cm]{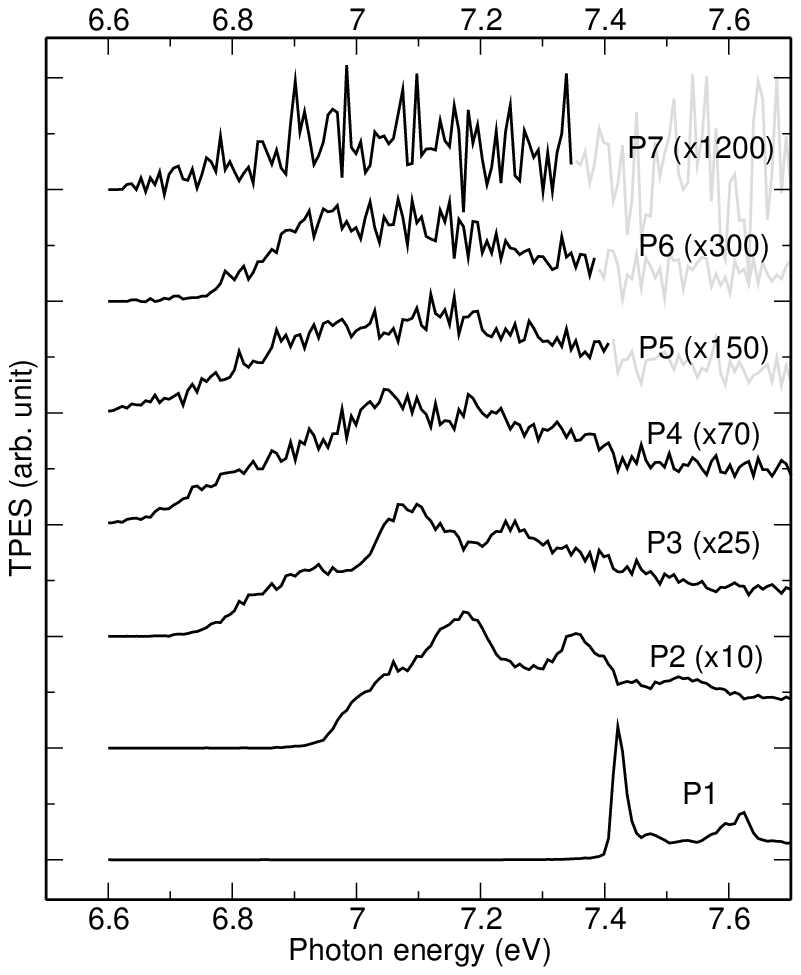} &
   \includegraphics[width=8.5cm]{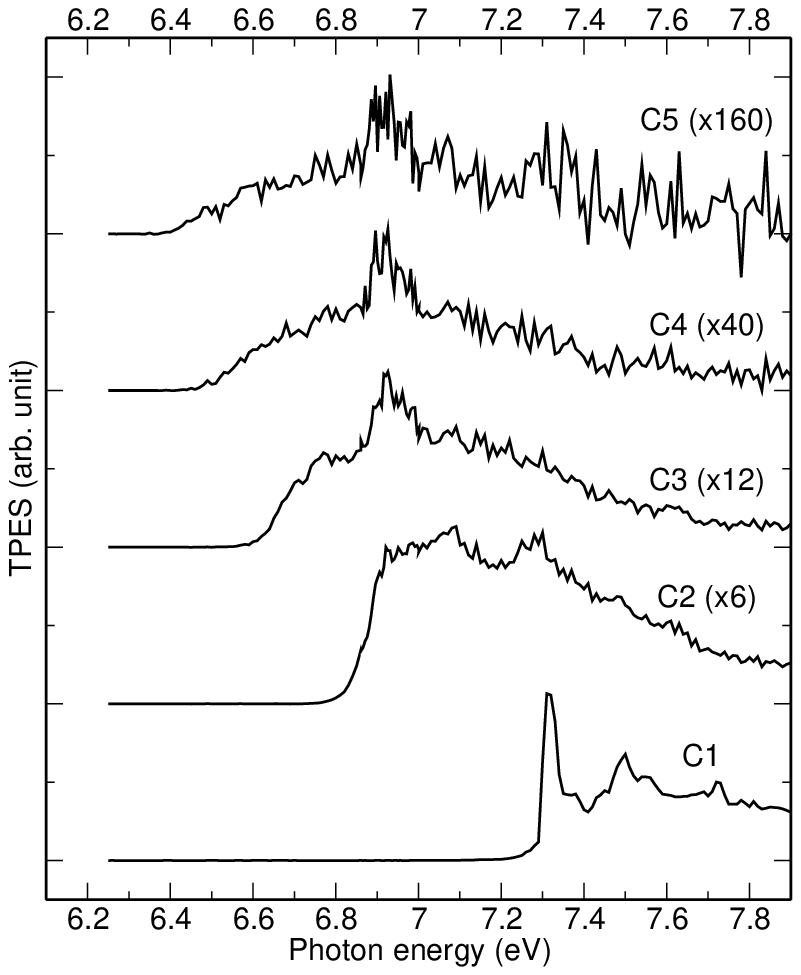}
  \end{tabular}
 \end{center}
\caption{Threshold photoelectron spectra (TPES) of the monomer and clusters of pyrene (P; left panel) and coronene (C; right panel), the labels P$_{i=1-7}$ and C$_{j=1-5}$ refering to the number of monomer units. The overall energy resolutions achieved were 15 meV and 30 meV for coronene and pyrene clusters, respectively.}\label{fig_tpes}
\end{figure*}

\begin{figure*}[ht]
 \begin{center}
  \begin{tabular}{cc}
   \includegraphics[width=8.5cm]{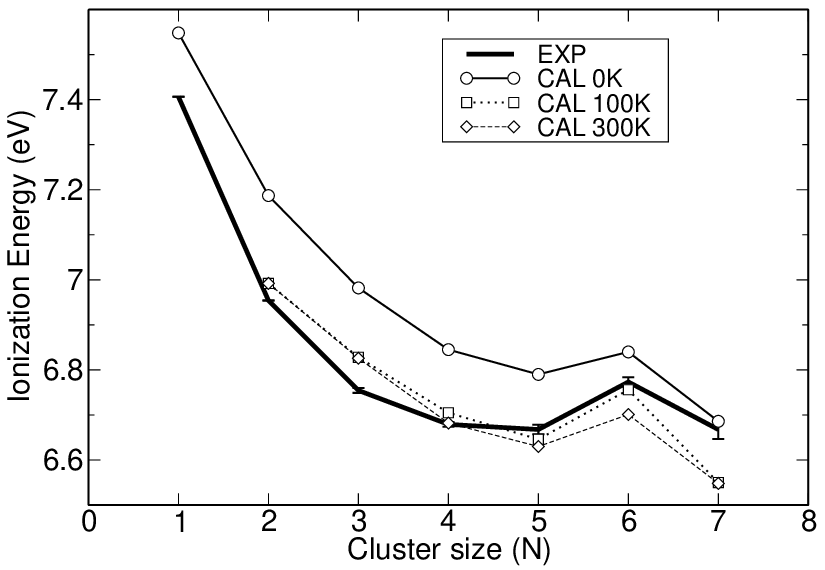} &
   \includegraphics[width=8.5cm]{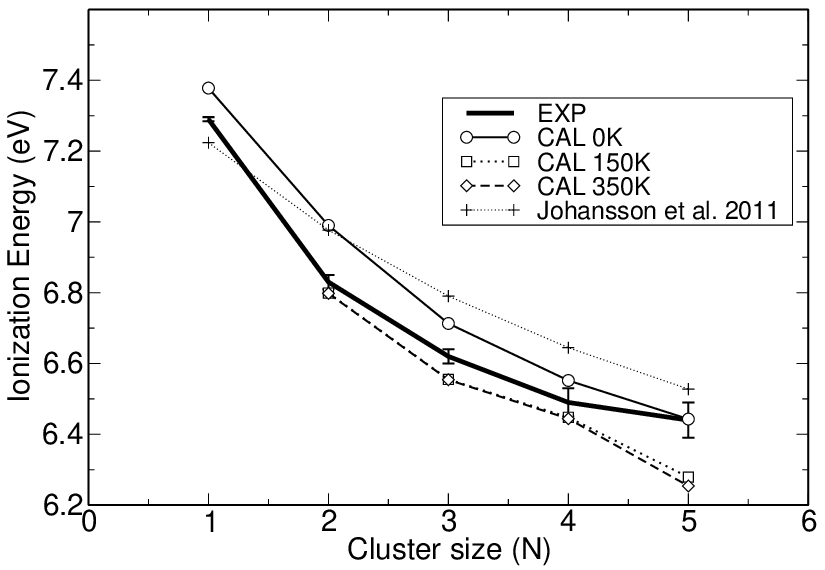}
  \end{tabular}
 \end{center}
\caption{Evolution of the ionization energy with cluster size for pyrene (left) and coronene (right) clusters. Experimental values are compared with calculated values at 0\,K and two finite temperatures. The values from the size scaling formula of Johansson et al. \cite{johansson11} are also shown. The one-sigma experimental error bars only take into account the statistical error, and are estimated assuming a Poisson distribution on the electron image pixels.}\label{fig_ip}
\end{figure*}

\begin{figure*}[ht]
 \begin{tabular}{cccccc}
   \includegraphics[width=2.3cm]{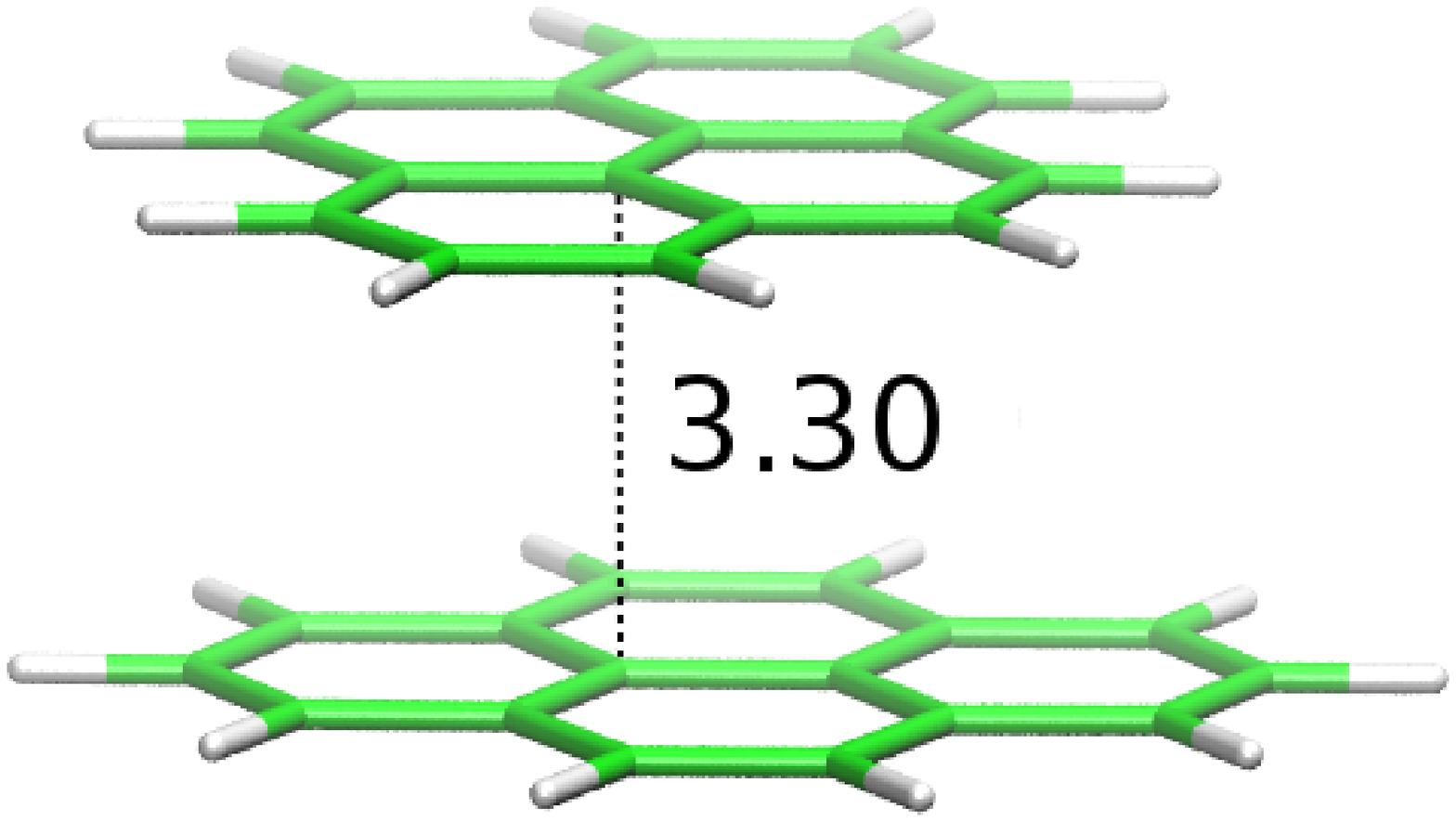} & 
   \includegraphics[width=2.3cm]{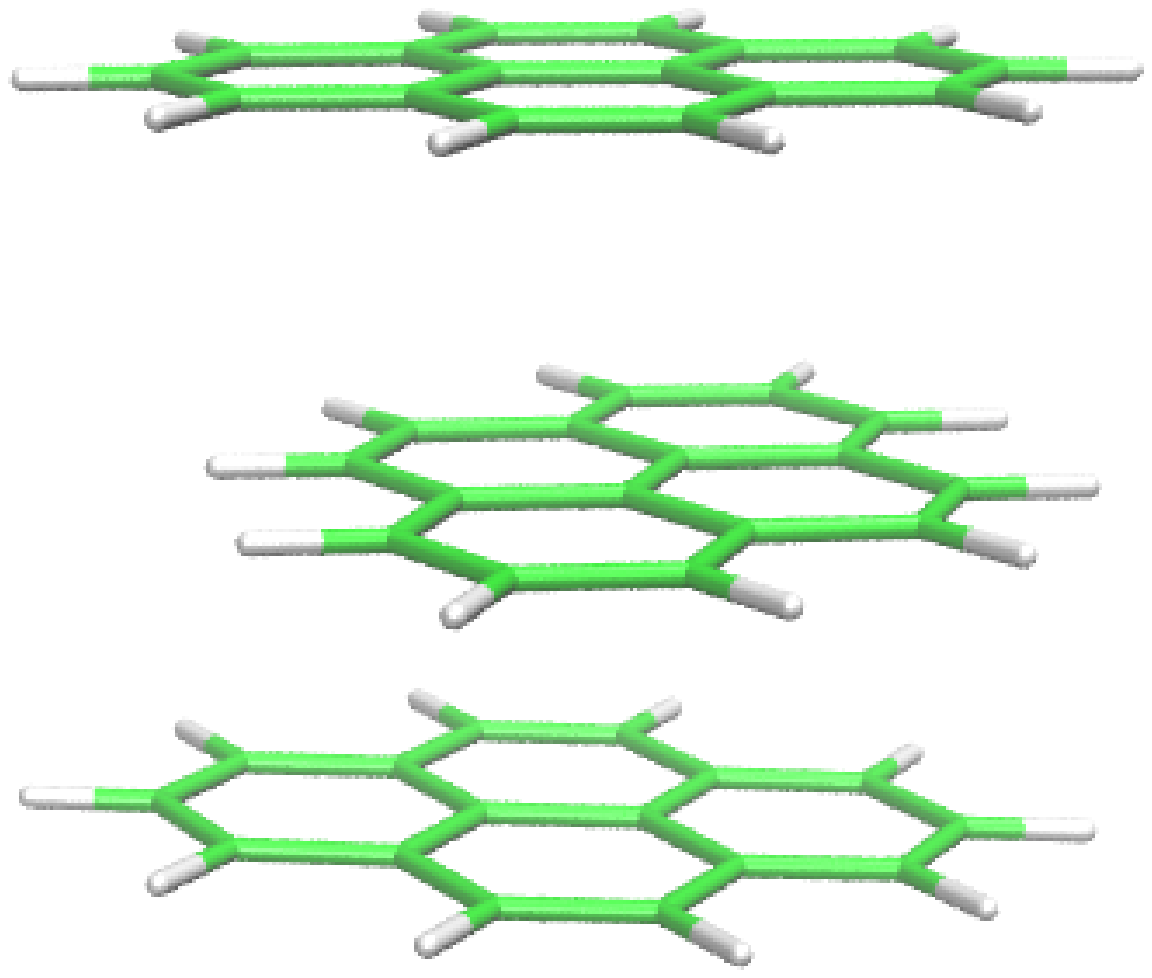} &
   \includegraphics[width=2.7cm]{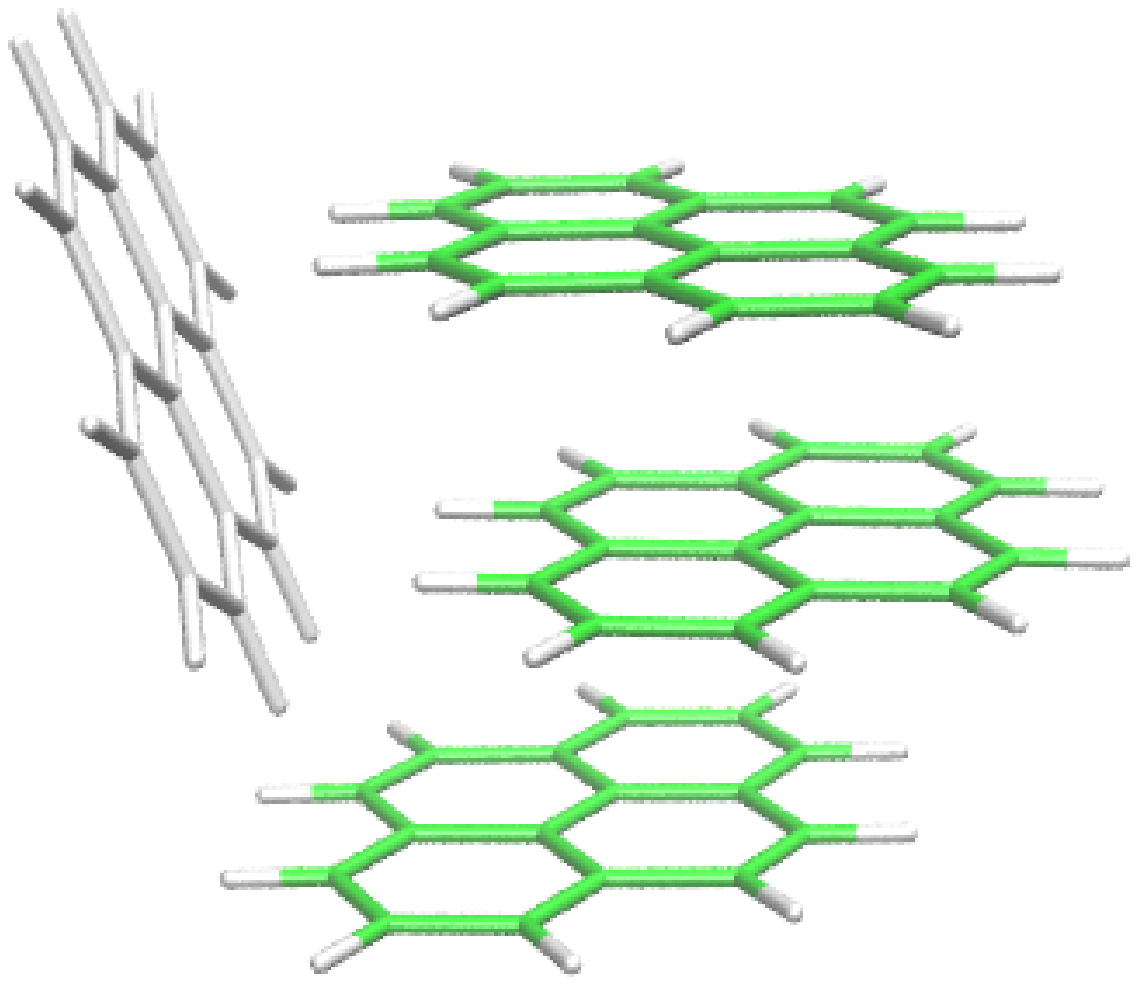} &
   \includegraphics[width=2.7cm]{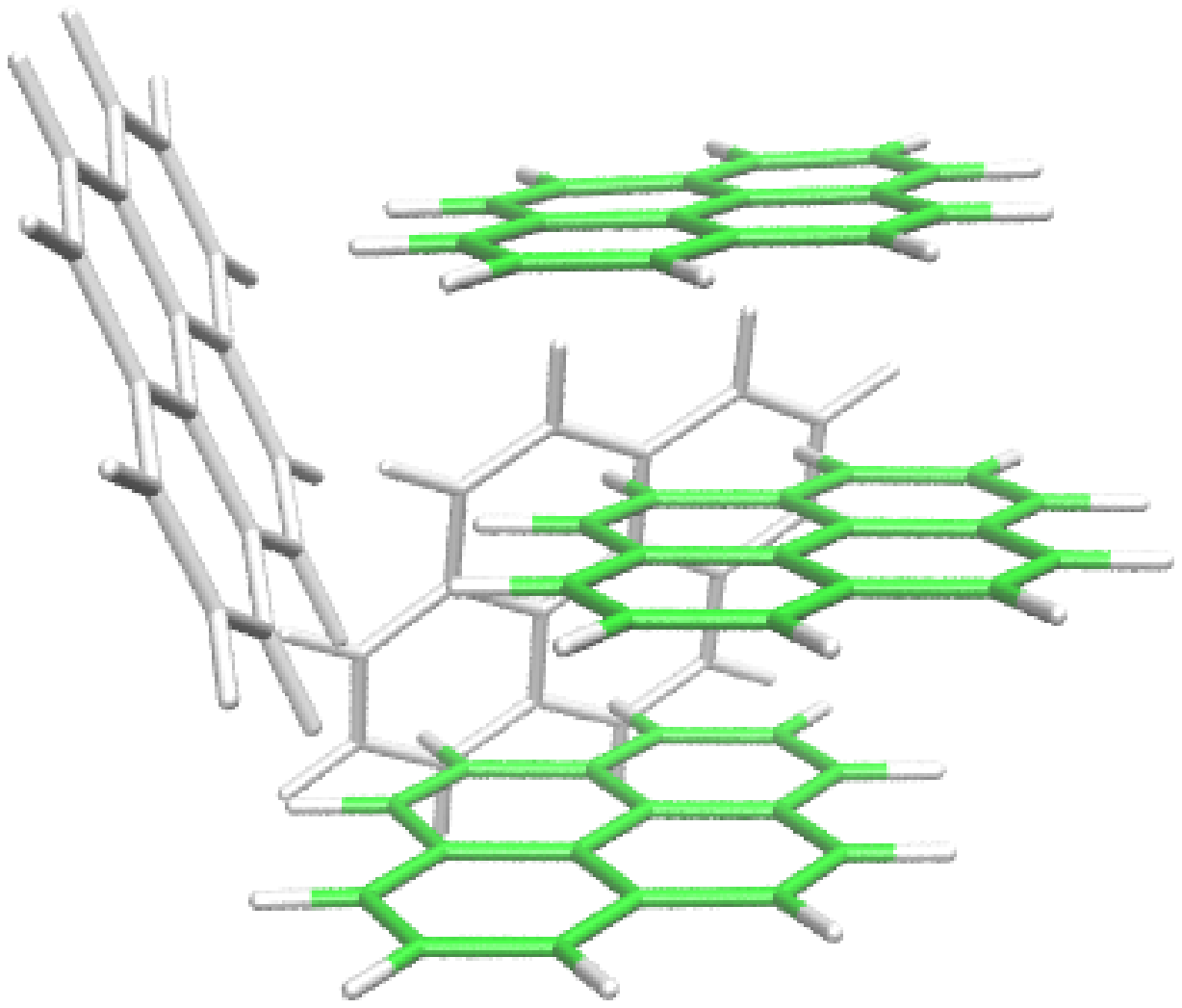} &
   \includegraphics[width=2.9cm]{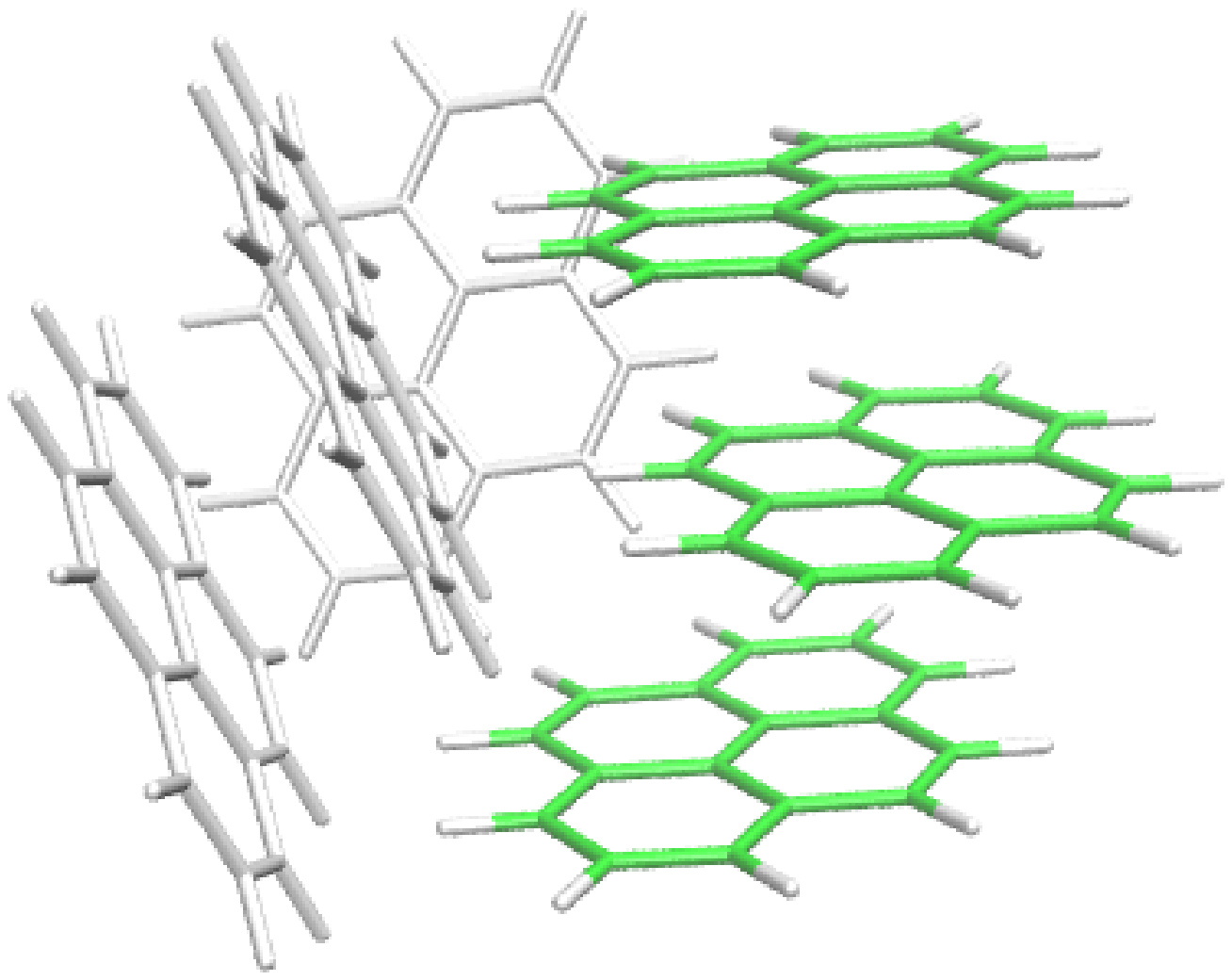} &
   \includegraphics[width=2.9cm]{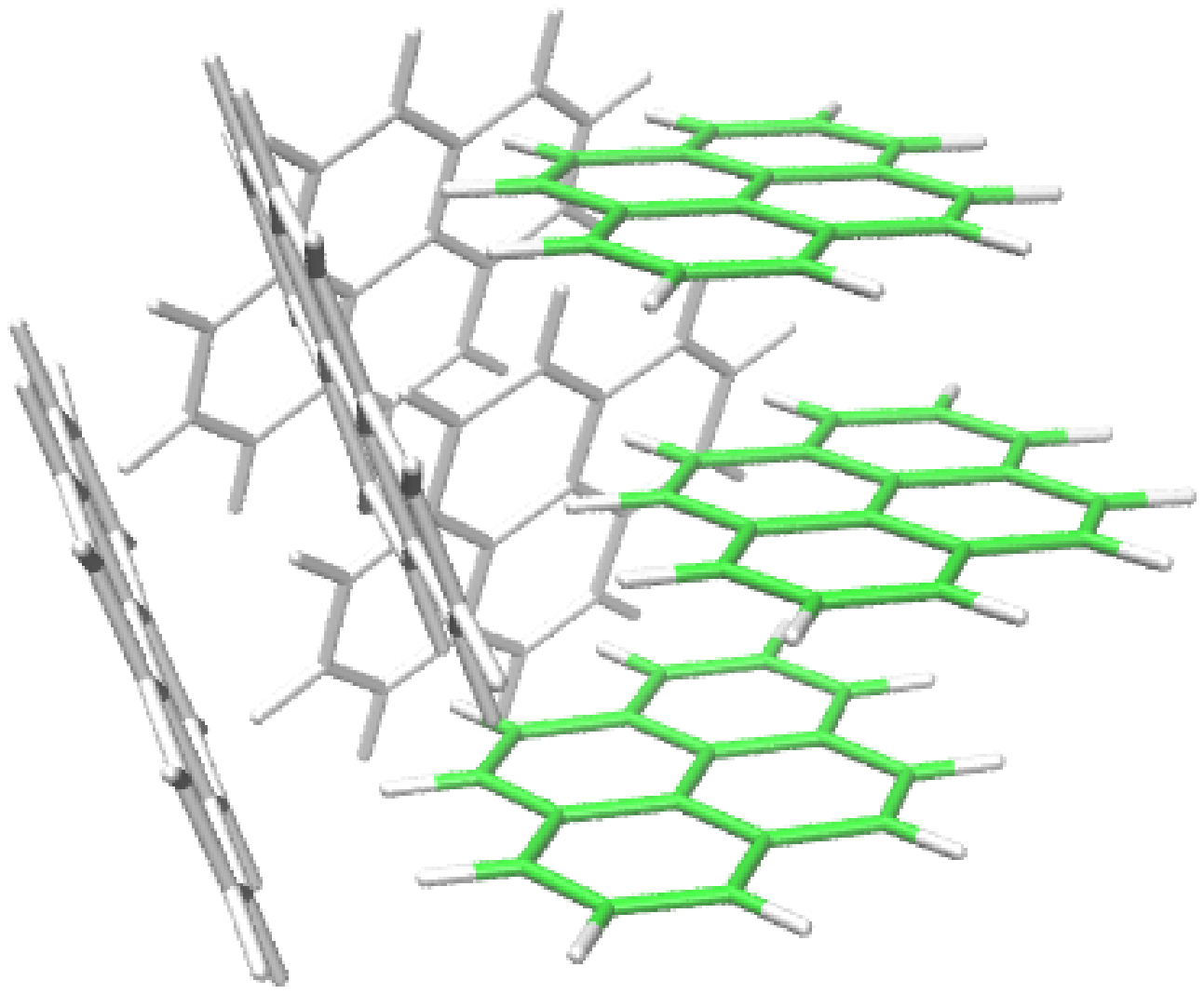} 
 \end{tabular}
 \vspace{-0.5cm}
  \begin{tabular}{cccc}
   \includegraphics[width=2.3cm]{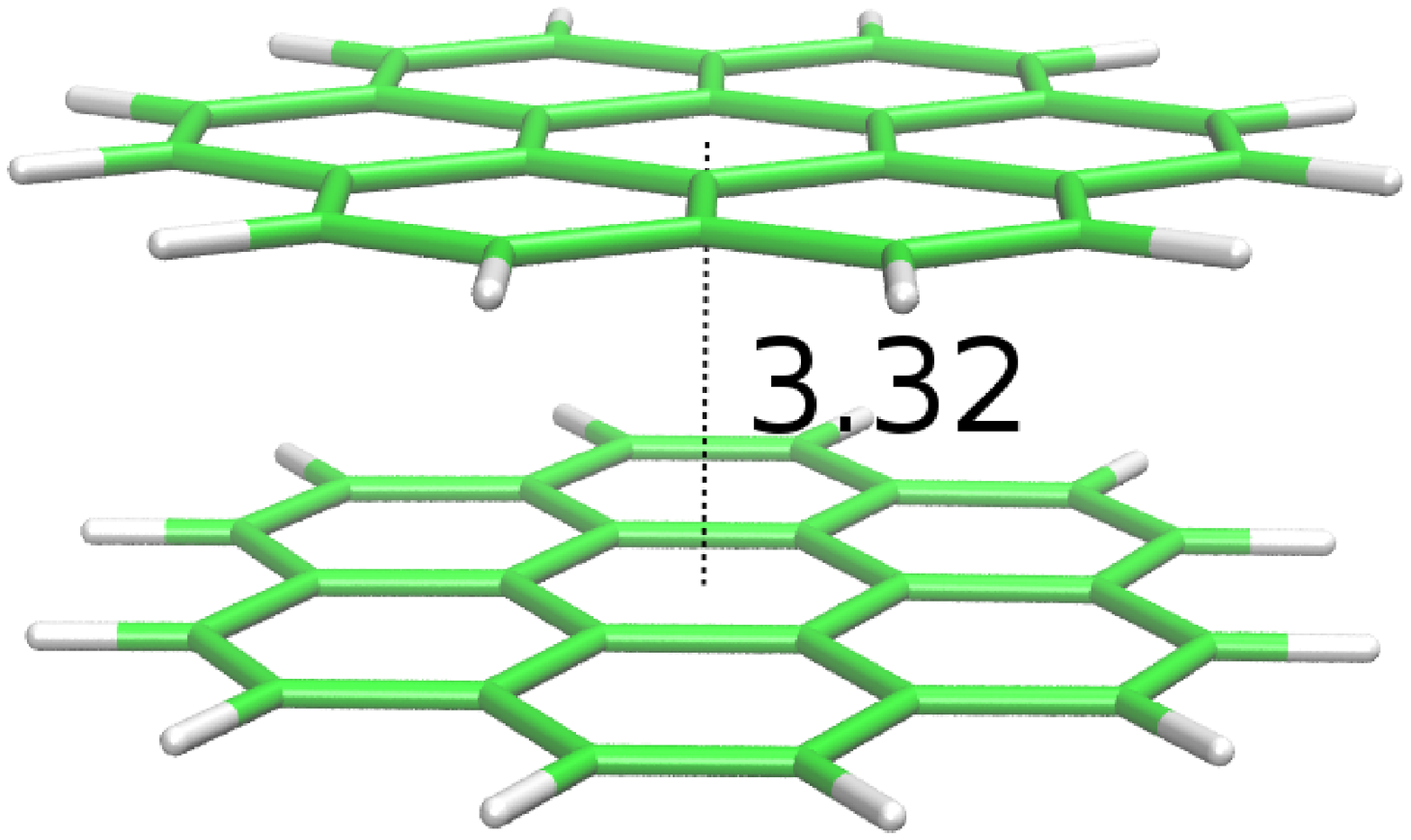} & 
   \includegraphics[width=2.5cm]{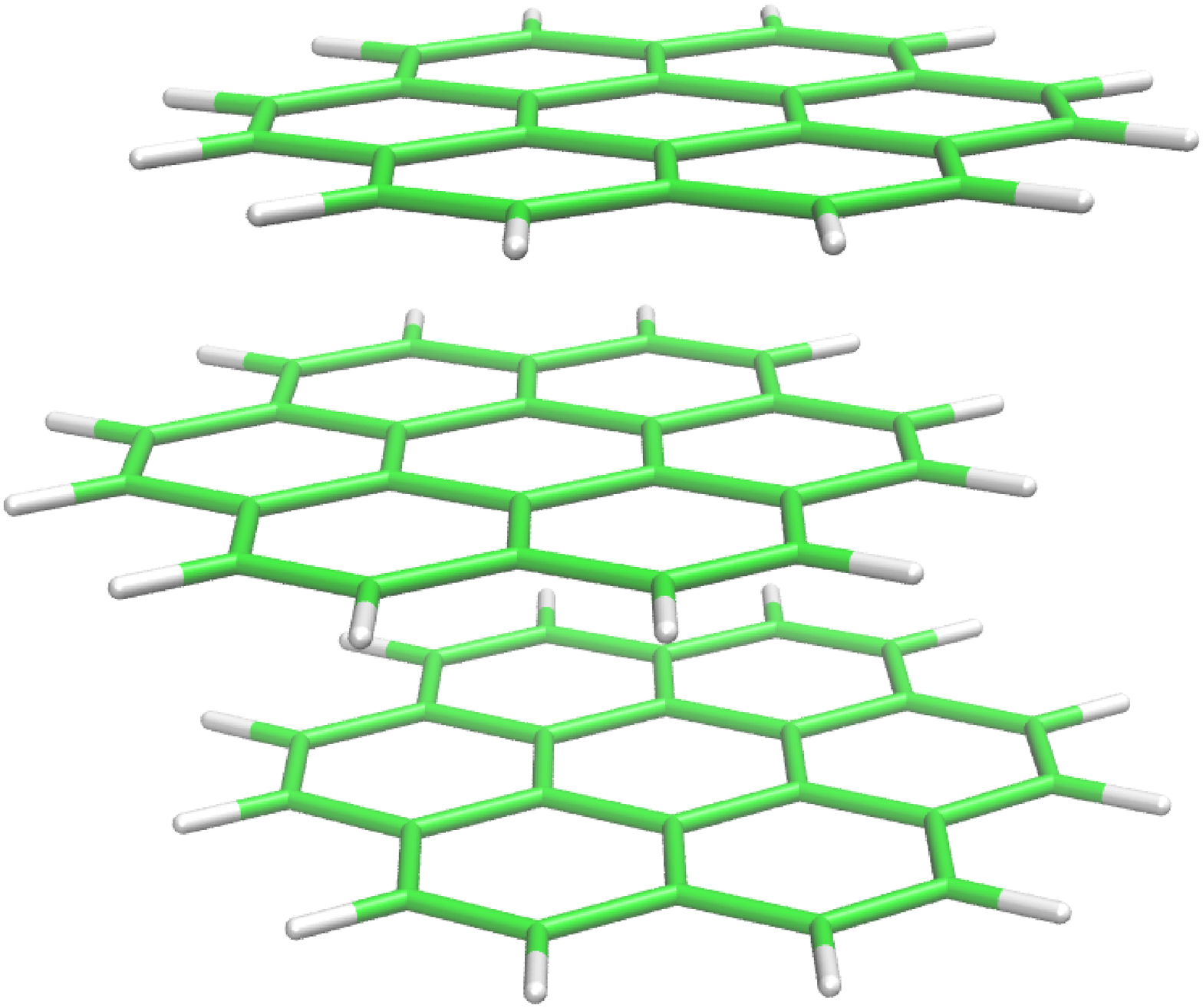} &
   \includegraphics[width=2.5cm]{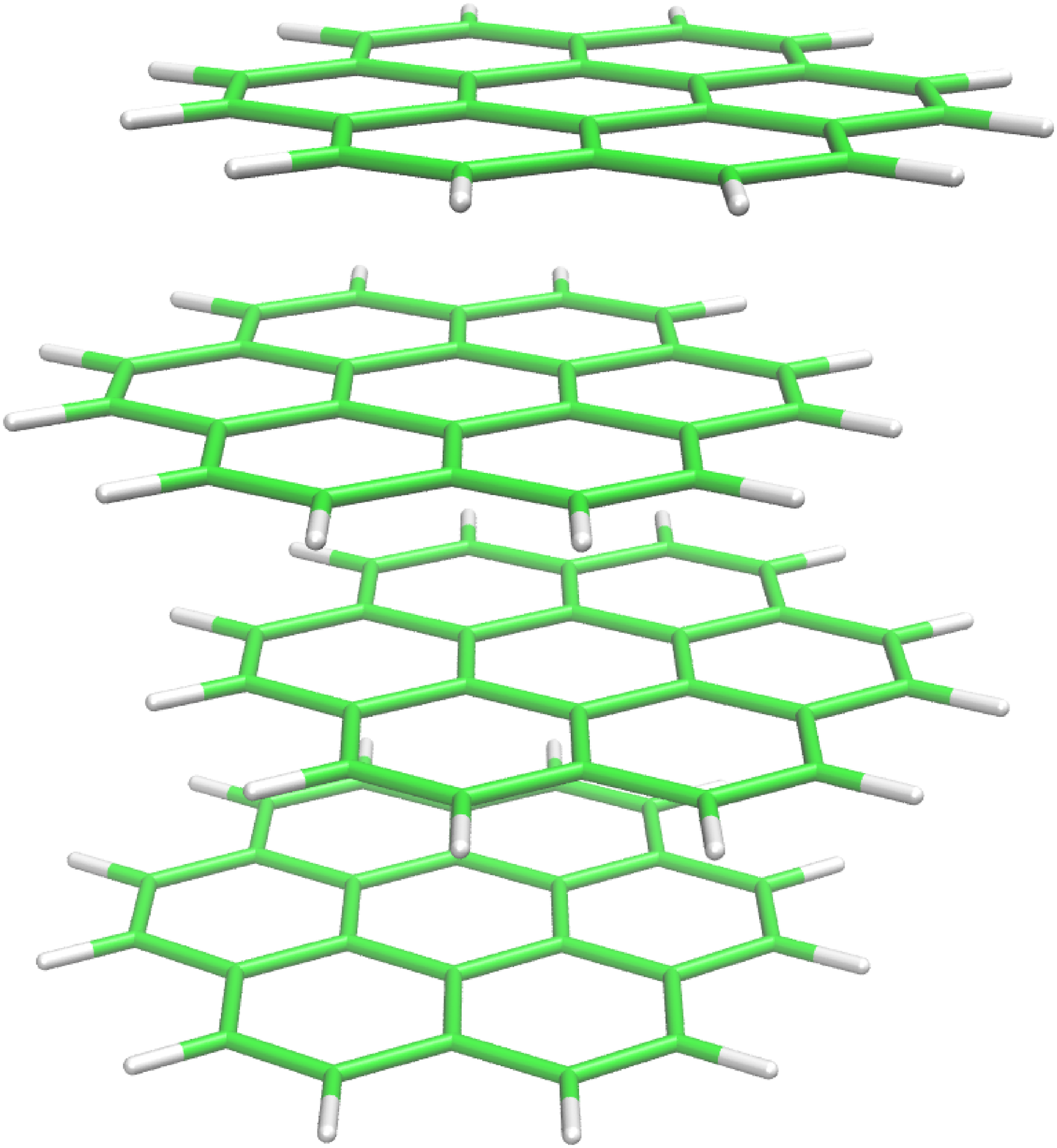} &
   \includegraphics[width=3.25cm]{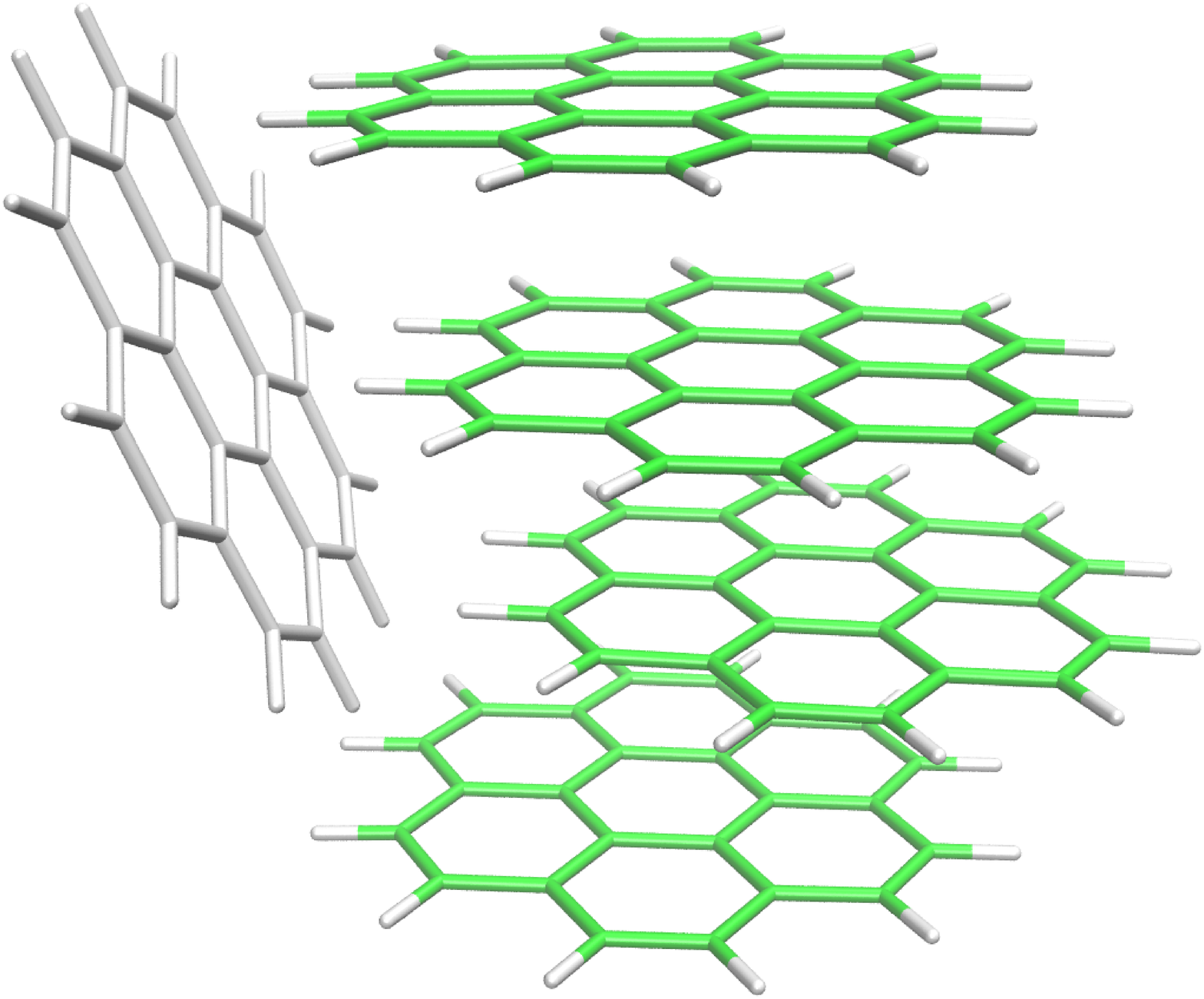}
 \end{tabular}
 \caption{Neutral structures for pyrene (upper row) and coronene (lower row) clusters. In cationic states, green molecules carry the charge and white molecules are mostly neutral. The indicated distances are provided in \AA. }\label{geom_neutral}

\end{figure*}

The mass spectrum displayed in Figure \ref{fig_tof} presents the relative intensity distribution of the pyrene clusters. It confirms that they are observed up to the heptamer at m/z=1414, and that no fragmentation process occurs in the studied energy range. The recorded threshold photoelectron spectra (TPES) are shown in Fig. \ref{fig_tpes} for both pyrene and coronene clusters up to an energy of 7.9\,eV. When increasing the size of the clusters,  both a  broadening of the spectral features and a  shift of the ionization onset towards lower energies are observed. The noise significantly increases for the largest clusters due to their low abundance. In the following, we focus solely on the  ionization energy (IE). Considering the complex shape of these curves, we chose to arbitrarily define IE for each species as the lowest value of the photon energy for which the TPES signal reaches 10\% of the local maximum. The evolution of these IE values with cluster size is displayed in Fig. \ref{fig_ip}.

From the theoretical side, the electronic structure of the clusters and potential energy surfaces are described using self-consistent Density Functional based Tight Binding methodology (DFTB)\cite{Elstner98} with dispersion corrections \cite{DFTB_CM3}. In order to ensure a proper description of the charge delocalization in ionized clusters, we used a scheme combining Valence Bond type Configuration Interaction and Constrained  DFTB \cite{rapacioli11, dontot16}. Structural optimization was performed using a global search algorithm. Parallel-tempering Monte-Carlo (MC) simulations  \cite{calvoptmc} involving configurational exchanges between temperature adjacent trajectories,  were first conducted in the  rigid molecule approximation to sample low-energy configurations. Each of these configurations was then further optimized via conjugated gradient scheme within an all-atom relaxation (namely intra- and inter-molecular degrees of freedom).

Figure\,\ref{geom_neutral} reports the lowest-energy structures that have been derived from this study. As already exemplified in previous works using simpler potentials \cite{rapacioli06}, one can see that the single stack geometry only prevails for the lowest sizes, up to 3 and 4 for pyrene and coronene, respectively. When increasing the size, multiple-stack structures become more favorable. From these structures, an IE can be determined assuming vertical transitions as stated in the Franck-Condon approximation, i.e. the energies for ionized clusters are calculated at the neutral geometry. This corresponds to the 0\,K value shown in Fig.\,\ref{fig_ip}. Note that in the case of coronene clusters, the present T=0K CI-DFTB results  (for globally optimized clusters) are in close correspondence with previous DFTB-CI results \cite{rapacioli09} restricted to regular stack geometries and confirmed by Johansson et al. with DFT \cite{johansson11}.  We also report on Fig.\,\ref{fig_ip} the IE values from the size scaling formula given in the latter reference.

At finite temperatures, a vertical IE is determined for all isomers and a theoretical ionization spectrum (IS) is built by assigning to each isomer (i) a Gaussian energy-distribution associated with the entropic motion within its potential energy surface basin and (ii) a Boltzmann weight with respect to its energy E$_i$ above the energy E$_0$ of the most stable isomer. This leads to:
\begin{equation}
 \text{IS}(E)=N \sum_i^{\text{iso}} \exp{\left[-\frac{E_i - E_0}{kT}\right]}\exp{\left[-\frac{(E-E_i)^2}{2\sigma^2}\right]}
\end{equation}
where $N$ is a normalization factor and $\sigma$ the standard deviation of the Gaussian related to the FWHM. The latter was set to 0.15 eV for pyrene and 0.2 eV for coronene, estimated from MC simulations on the dimer species.
A theoretical estimation of the appearance IE can be extracted from the simulated spectra using the same 10\% level procedure as that performed for the experimental data. This scheme was repeated at  T=100\,K and 300\,K for pyrene clusters and at T=150 and 350\,K for coronene clusters. These values encompass the range of temperatures expected  for the clusters in the experiment. Note that we have previously determined a typical vibrational temperature of 200\,K for the coronene monomer in the molecular beam \cite{brechignac14}.

As shown in Fig. \ref{fig_ip},  all theoretical curves show a general trend in very good agreement with the experimental values. Inclusion of temperature effects leads to a global decrease by 0.1-0.2\,eV relative to the values at 0\,K. In the considered range of temperature the major temperature effect is seen on the hexamer of pyrene with a shift close to 0.1 eV between 100 and 300\,K.  For both the hexamer (P6) and the heptamer (P7), 4 isomers represent 95\% of the thermal population at 100\,K, whereas 22 isomers are involved at 300\,K. The stronger temperature effect on P6 is related to the range of vertical IE values associated to these isomers. This range is much larger for P6: 0.66 eV (0.22 eV) at 300\,K (100\,K) compared to 0.24 eV (0.09 eV) for P7 . While the energy of the neutral isomers are quite close, the energies of the ions at those 
geometries may differ significantly. The bump observed in the theoretical IE for P6 is due to the fact that, at the geometry of the neutral, this ion is less stable relative to its neighbours.
Finally, we note that the IEs of the largest clusters seem to be better matched by the 0\,K model for both pyrene and coronene, whereas small clusters are better explained by the models at temperatures higher than 100\,K. 
Although very tentative, this  observation suggests that higher-mass clusters are at lower temperature than lower-mass clusters, in line with the expectations from the "evaporative ensemble" model, first introduced to account for the relative intensities of alkali clusters \cite{cbrechignac89}. One should however keep in mind that the temperature effects could be of the same magnitude as the errors of the DFTB-CI evaluation of the energy and that of anharmonic effects, which are not included here.

This study reports the first experimental determination of the IE of PAH clusters in the range N=2-7.  The IE is found to decrease with cluster size, with the exception of the pyrene hexamer. The consistency between experiment and theory gives support to the geometries shown in Fig. 4 and to the ability of currently developed models to describe the electronic properties of these large systems even when charge delocalization and resonance have to be taken into account. 

In future works, the model could  also be used beyond the ionization onset for the analysis of the electronic structure revealed in the TPES (Fig.\,\ref{fig_tpes}) in order to help elucidating the contribution of isomers versus vibrational structure above threshold. Since thermalization is likely not achieved in our experiment, a detailed comparison with the model would however benefit from experimental data on thermalized clusters.
Finally, as a first implication for astrochemistry,  this study shows that PAH clusters can be ionized at significantly lower energies than the monomers. 
This result, combined with the expected higher ionization cross-section for these species, suggests that ionized PAH clusters are likely to be formed in astrophysical PDRs, making it worth exploring their survival in these environments.

\begin{acknowledgments}
We acknowledge the financial support of the Agence Nationale de la Recherche through the GASPARIM project �Gas-phase PAH research for the interstellar medium� (ANR-10-BLAN-0501) and computing ressources by CALMIP supercomputing center (allocation 2015-P0059). L.D. thanks the ERC for support under grant ERC-2013- Syg-610256-NANOCOSMOS.
\end{acknowledgments}

\bibliography{biblio}

\end{document}